\documentclass[twocolumn,aps,showpacs,floatfix]{revtex4}

\usepackage{graphicx}
\usepackage{bm}
\begin{document}

\title{Enhancement of the scissors mode of an expanding Bose-Einstein
 condensate}

\author{M. Modugno}
\author{G. Modugno}
\author{G. Roati}
\altaffiliation[Also at: ]{Dipartimento di Fisica, Universit\`a di
    Trento, 38050 Povo, Italy.}
\author{C. Fort}
\author{M. Inguscio}
 \affiliation{%
 INFM - LENS - Dipartimento di Fisica, Universit\`a di Firenze\\
 Via Nello Carrara 1, 50019 Sesto Fiorentino, Italy }%

\date{\today}

\begin{abstract}
We study the time-evolution of the scissors mode of a
Bose-Einstein condensate during the ballistic expansion after
release from the magnetic trap.  We show that despite the
nontrivial character of the superfluid expansion, the sinusoidal
behavior of the scissor oscillations is recovered after an
asymptotic expansion, with an enhancement of the final amplitude.
We investigate this phenomenon with a condensate held in an
elongated magnetostatic potential, whose particular shape allows
for the excitation of the scissors mode.
\end{abstract}


\maketitle

The close connection between Bose-Einstein condensation in dilute
atomic gases and superfluidity has been shown by recent research.
The most striking signatures of superfluidity observed are
quantized vortices \cite{vortici}, reduction of dissipative
phenomena \cite{ketterle,firenze}, scissors modes
\cite{scissors,scissors_foot}, and irrotational flow
\cite{edwards,foot}.

Due to the particular requirements for the trapping potential, the
observation of the scissors mode has been so far reported only for
disk shaped condensates \cite{scissors_foot}, where the superfluid
behavior is evidenced by the different oscillation frequencies with
respect to the thermal gas \cite{scissors}. Very recently, the authors
of \cite{edwards,foot} have also analyzed the implication of
irrotationality on the expansion of Bose-Einstein condensates (BECs)
initially confined in a rotating trap, showing that the condensate expands 
in a distinctively different way with
respect to a non-rotating condensate, thus providing a direct
manifestation of the quenching of the moment of inertia.

In this paper we show that the peculiar behavior of a free-expanding
rotating condensate has also important consequences on the evolution of a
scissors mode initially excited in the trap. 
Although the condensate undergoes a nontrivial expansion, we find that
asymptotically the scissors mode recovers the initial 
sinusoidal behavior, with an enhancement of the final
amplitude.
To verify our predictions we take advantage of the particular shape of
the magnetic field in a magnetostatic trap, which permits the
excitations of scissors modes in elongated BECs.
This allows us to investigate the expansion of a rotating superfluid
in a novel regime, where the implications of the irrotationality are
the most dramatic.

The wave function of a condensate at zero temperature
can be conveniently written in terms of the density
$\rho$ and the phase $S$
\begin{equation}
\psi({\bf x}, t) = \sqrt{\rho({\bf x}, t)} e^{i S({\bf x}, t)/\hbar}.
\end{equation}
In the Thomas-Fermi approximation the Gross-Pitaevskii
equation for $\psi$ can be transformed in two coupled hydrodynamics
equations for the density $\rho$ and the velocity ${\bf v}=\nabla S/m$
\cite{stringari,BEC_review}
\begin{eqnarray}
\label{eq:hydro_rho}
&&\frac{\partial \rho}{\partial t} + {\bm\nabla}\left(\rho{\bf v}
\right)=0 \\ &&m\frac{\partial \bf v}{\partial t} +
{\bm\nabla}\left(\frac{{\bf v} ^2}{2 m} + U({\bf r}, t) + g
\rho\right)=0,
\label{eq:hydro_v}
\end{eqnarray}
where $U$ is the trapping potential, 
 $g=4\pi\hbar^2a/m$ the coupling strength, $m$ the
atomic mass and $a$ the inter-atomic scattering length.

Here we are interested in the solution of Eqs.\
(\ref{eq:hydro_rho})-(\ref{eq:hydro_v}) which corresponds to a scissors
mode, that is a shape-preserving oscillation of the condensate
\cite{scissors}. To account for this we consider a
harmonic  potential of the form 
\begin{equation}
U({\bf x}, t)\equiv \frac{1}{2}m\omega_{ho}^2\sum_{ij=1}^3
x_iW_{ij}(t)x_j
\label{eq:potential}
\end{equation}
where $W=W^T$ for symmetry reasons, 
with initial conditions 
$W_{ij}(t<0)=\omega^2_i\delta_{ij}/\omega^2_{ho}$, and
$\omega_{ho}\equiv(\omega_x\omega_y\omega_z)^{1/3}$.
By tilting the trap potential by a small angle
$\theta_{0i}$ around the $z$-axis at $t=0$, one can excite
a scissors mode in the $x-y$ plane,
identified by the angle
\begin{equation}
\theta_0(t)=\theta_{0i}\cos(\omega_{sc}t),
\label{eq:scissor}
\end{equation}
whose frequency is $\omega_{sc}=(\omega_x^2+\omega_y^2)^{1/2}$ \cite{scissors}.
In this case Eqs.\ (\ref{eq:hydro_rho})-(\ref{eq:hydro_v}) can
be solved exactly with a quadratic ansatz for the condensate density and
phase
\cite{BEC_review,tf}
\begin{eqnarray}
\label{eq:ansatz_rho}
&&\rho({\bf x}, t) = {m\omega_{ho}^2\over g}\left(\rho_0(t)
-{1\over2}\sum_{i,j =
1}^{3} x_{i} A_{ij}(t) x_{j}\right),\\
&&S({\bf x}, t) =
{m\omega_{ho}}\left(s_{0}(t) + {1\over2}\sum_{i,j = 1}^{3}x_{i}
B_{ij}(t) x_{j}\right).
\label{eq:ansatz_s}
\end{eqnarray}
Therefore, in general the system can be described by 10 time dependent
dimensionless parameters: $\rho_0$, $S_0$, $A_{ij}$ and $B_{ij}$ ($A$
and $B$ are $3 \times 3$ symmetric matrices, reflecting the symmetry
property of $U$, with $A_{i3}=0=B_{i3}$ due to the vanishing of
$W_{i3}(t)$).  These parameters obey the
following set of first order differential equations \cite{tf}
\begin{eqnarray}
\frac{d\rho_0}{d\tau} & = & \rho_0\mbox{Tr} B \,;\,\, \frac{d
S_0}{d\tau} = \rho_0
\label{eq:rho_dot}\\
\frac{d A}{d\tau} & = & 2 A\mbox{Tr} B + \{A,B\}
\label{eq:A_dot}\\
\frac{d B}{d\tau} & = & W - A- B^2
\label{eq:B_dot}
\end{eqnarray}
with boundary conditions determined by the equilibrium distribution of
the condensate at $\tau\equiv\omega_{ho}t=0$.  
Notice that these equations depend on the
number of atoms $N$ and on the scattering length $a$ only through the
initial value of $\rho_0(0)=0.5a^2_{ho}(15Na/a_{ho})^{2/5}$.

The subsequent angular motion is evaluated in terms of the rotation
 which diagonalizes the quadratic part of the
density $\rho({\bf x},t)$, (that is the matrix $A_{ij}$). The angle
of rotation is fixed by the relation \cite{edwards}
\begin{equation}
\tan(2\theta_0)=\frac{2 A_{12}}{A_{11} -A_{22}}.
\end{equation}

The same Eqs.\ (\ref{eq:rho_dot})-(\ref{eq:B_dot}) can be used to study
the expansion of the condensate after the release from the trap ($W\equiv0$),
with the appropriate initial conditions.

Experimentally
we study the scissors mode of a BEC of $^{87}$Rb atoms in an elongated
potential created by a QUIC magnetostatic trap \cite{science}. The
potential has a cylindrical symmetry around the $x$-axis, with
frequencies $\omega_x$=2$\pi\times$16.3~Hz and
$\omega_y$=$\omega_z$=2$\pi\times$200~Hz. The BEC is produced by standard
radio-frequency evaporation, and is typically composed by 10$^5$
atoms. In all the measurements we report no thermal component was
detectable.

Since the trap is static, the scissors mode cannot be
excited with the technique demonstrated for time-orbiting-potential
traps \cite{scissors_foot}.
However, we can take advantage of the fact that 
gravity breaks the symmetry of the magnetic potential
resulting in anharmonicity even at the trap minimum.
Actually, we found that the eigenaxes of our magnetic trap 
are rotated in the $x-y$ plane in a localized region around
the minimum. The rapid motion of the condensate through such deformation
is therefore equivalent to a sudden rotation of the trap, 
like that considered in Eq.\ (\ref{eq:scissor}).
In fact, we found that when we excite a dipolar oscillation along the
weak $x$-axis by a sudden displacement of the trap minimum, a scissors mode
appears as soon as the BEC travels for the first time through the
minimum. Since for an elongated BEC the scissors mode proceeds
essentially at $\omega_y\gg\omega_x$, it is possible to study several
scissors oscillations before the BEC travels again through the center
of the trap. By choosing an appropriately small dipolar amplitude, we
observe a ``pure'' scissors mode with amplitude $\theta_{0i}\simeq15$~mrad, 
without any apparent shape deformation.

\begin{figure}
\centerline{\includegraphics[width=6.5cm,clip=]{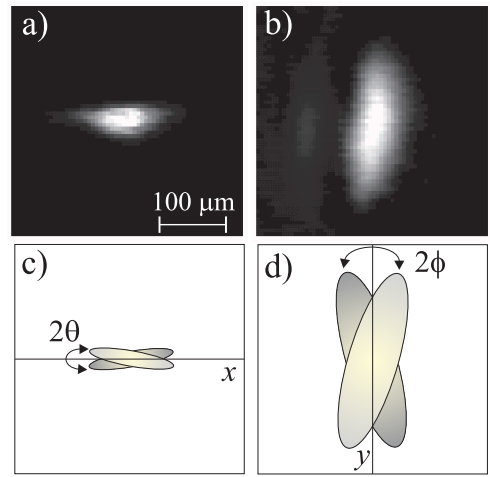}}
\caption{Absorption images of a rotating, elongated Rb condensate
during the ballistic expansion. In a) the BEC has just been
released from the trap ($t_{exp}=2$~ms), and in b) it has expanded
for 23~ms. The other two pictures, not to scale, show the
corresponding evolution of a scissors mode. The initial
oscillation about the horizontal $x$-axis (c) is transformed in an
oscillation about the vertical $y$-axis for long expansion times
(d).} \label{fig:image}
\end{figure}

To describe the characterization of such mode, we need
first to recall the peculiar behavior of a rotating BEC during the
expansion after release from the trap \cite{edwards}. In fact, due
to the irrotational nature of a superfluid ($\nabla\times{\bf 
v}=0$) the moment of inertia $\Theta$ of a condensate is quenched
with respect to the rigid body value $\Theta_{rig}$ \cite{zambelli}
\begin{equation}
\Theta=\frac{\langle x^2-y^2\rangle^2}{\langle
x^2+y^2\rangle^2}\Theta_{rig}.
\label{eq:inertia}
\end{equation}
Therefore, due to conservation of energy and angular momentum, a
rotating condensate cannot reach a symmetric configuration during
the expansion, and undergoes a rapid rotation causing the
inversion of the ``long'' and ``short'' axes. In particular, if at
release the condensate is rotating in the $x-y$ plane, with its
long axis forming a small angle $\theta_0$ with the $x-$axis, as
shown in Fig.\ \ref{fig:image}, initially it will start expanding
in the short direction. Then, when the aspect ratio approaches
unity, the condensate will continue to expand in the long
direction, with its long axis rotated by an angle which is close
to but not exactly $\pi/2$ \cite{scissors}.

\begin{figure}
\centerline{\includegraphics[width=7.5cm,clip=]{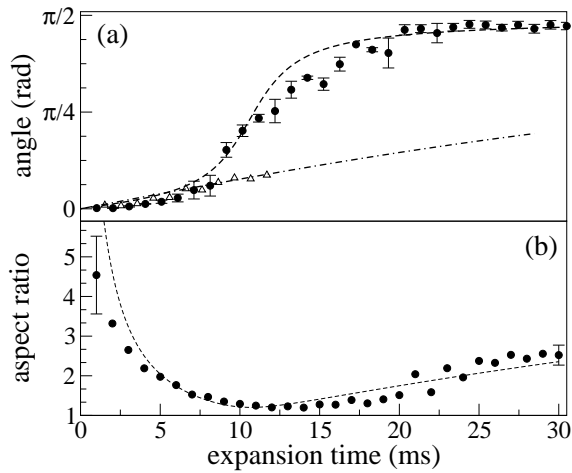}}
\caption{a) Evolution of the rotation angle for a condensate
(circles) and a thermal cloud (triangles) released with small angular
velocity from an elongated trap. b) Evolution of the aspect ratio of the
rotating condensate. When the aspect ratio gets close to unity
 the angular velocity shows a rapid increase.
The lines represent the theoretical predictions.}
\label{fig:scissexp}
\end{figure}

We have studied the evolution of the rotation angle $\theta$ of the
condensate as a function of the expansion time $t_{exp}$, as shown in
Fig.\ \ref{fig:scissexp}(a). The measurements are performed by
releasing the condensate after a fixed evolution time $t$ of the
scissors oscillations in the trap. For comparison we also plot the
angle evolution of the thermal cloud just above condensation, 
which has been put under rotation with the same technique described above.
Note that the rotation of the thermal cloud can be experimentally
investigated only until the cloud becomes spherical at 
$t_{exp}\simeq1/\omega_x$.
Initially, the expansion of the BEC and the thermal cloud 
is almost indistinguishable,
being characterized by the angular velocity $\Omega_0$ at release
from the trap. This is due to the fact that the moment of inertia of an 
elongated condensate is close to the rigid body value 
(see Eq.\ (\ref{eq:inertia})). The quenching becomes important 
as soon as the aspect ratio approaches unity, where the 
non-classical behavior of the condensate is evident, as predicted in 
\cite{edwards}. Indeed, its
rotation undergoes a fast acceleration (Fig.\
\ref{fig:scissexp}(b)), followed by a slow evolution towards an
asymptotic angle, close but smaller than $\pi$/2. The agreement of
experiment and theory  is quite good.
In contrast, the rotation angle of the thermal cloud continues to
follow the predicted behavior \cite{edwards}
\begin{equation}
\theta(t;t_{exp})=\theta_0(t)+ \mbox{atan}(\Omega_0(t) t_{exp})\,,
\label{eq:thermal}
\end{equation}
as shown in Fig.\ \ref{fig:scissexp}(a).  This figure shows that the use
of very elongated condensates allows for the investigation of novel regimes 
where the superfluid expansion is dramatically different from the
classical one.

\begin{figure}
\centerline{\includegraphics[width=7.5cm,clip=]{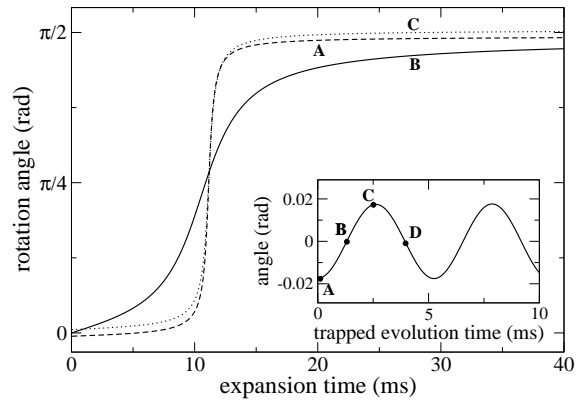}}
\caption{Calculated rotation angle of the condensate during the
expansion, released at different times during the scissors oscillation
in the trap, as shown in the inset. Cases A and C correspond to small
angular velocities, B and D to the maximum velocity. The evolution of
D is the opposite of B (negative angles).}
\label{fig:scissexp-th}
\end{figure}

By analyzing the expansion of the condensate for the various
conditions of initial angle and angular velocity at release from the
trap, it is possible to reconstruct the evolution of a scissors
oscillation for the expanding BEC. As an example, in
Fig.\ \ref{fig:scissexp-th} we show the calculated evolution of the
angle $\theta$ during the time of flight, for conditions
corresponding to very small angular velocity  and maximal angular 
velocity of the trapped oscillations.  
The behavior of $\theta$ in all these regimes is
qualitatively similar to that shown in
Fig.\ \ref{fig:scissexp}. However, in the situations of small initial
angular velocity the transient rotation is even faster than
in the other cases, while the asymptotic angle is closer to $\pi$/2.

By studying the solutions of Eq.\ (\ref{eq:rho_dot})-(\ref{eq:B_dot}),
we can identify three different regimes for the expanding scissors mode, 
depending on the
expansion time $t_{exp}$. i) For $t_{exp}<1/\omega_x$ the condensate expands
in a way similar to a classical gas, and the sinusoidal behavior of
the scissors mode is preserved, as is possible to see from
Eq.\ (\ref{eq:thermal}). ii) When $t_{exp}\simeq1/\omega_x$ the rotation
angle is always close to $\pm\pi/4$, depending on the sign of the angular
velocity at release. The angle oscillation is therefore close to
a square wave with the same phase of the angular velocity in the trap
$\Omega_0(t)$. iii) For $t_{exp}>1/\omega_x$ the angle of the condensate
approaches $\pi/2$, and it is more convenient to study the
oscillations of the angle $\phi$ from vertical. We find that the
original sinusoidal behavior of the oscillations is restored, with a
time-dependence which can be fitted by the form
\begin{eqnarray}
\label{eq:fit}
\phi(t;t_{exp})&=&\theta_0(t)+F(t_{exp})\Omega_0(t)\\
&=&
\theta_{0i}\sqrt{1+\omega_{sc}^2 F^2(t_{exp})}\cos(\omega_{sc}t+\varphi)
\nonumber\,,
\end{eqnarray}
where $F$ is a non-trivial function of the expansion time and of the
trap geometry, and $\varphi$ is a phase shift.
This behavior is remarkable, since
in passing over the ``critical'' time-region at $t_{exp}\simeq
1/\omega_x$, the condensate does not lose memory of its initial
angular velocity, and it starts to behave again as a classical gas.
\begin{figure}
\centerline{\includegraphics[width=7.5cm,clip=]{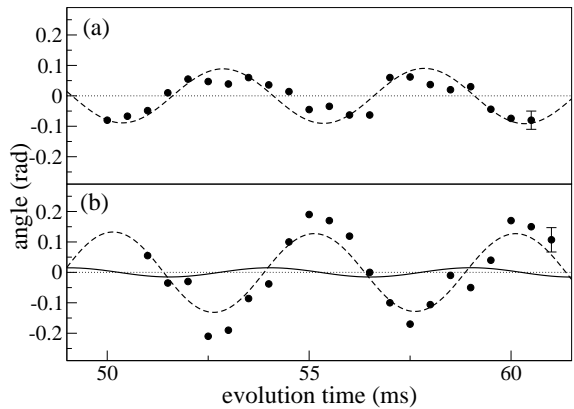}}
\caption{Evolution of the scissors mode during the ballistic
expansion. In a) the angle oscillation $\theta(t)$ after a short
expansion of 4~ms (circles) is compared to theory (dashed line) to
extrapolate the oscillation in the trap.  In b) we compare the latter
(continuous line) to the oscillation $\phi(t)$ after a longer
expansion of 23~ms (circles). The data show an enhancement of 
the scissors mode, as expected from theory (dashed line).}
\label{fig:amp-exp}
\end{figure}

We have been able to verify experimentally these expectations in
our Rb BEC. Since the finite resolution of the imaging system
prevents us from studying the condensate inside the trap, we probe
the oscillation of the condensate after a minimum expansion
$t_{exp}=4$~ms, as shown in Fig.\ \ref{fig:amp-exp}(a), and we
reconstruct the scissors mode in the trap. In Fig.\
\ref{fig:amp-exp}(b) we compare such a motion to
the experimental scissors mode after a long expansion of 23~ms. 
The initial scissors mode is clearly amplified and
phase-shifted after the expansion. For the experimental parameters
we calculate indeed that $\omega_{sc}F(t_{exp}=23~ms)\simeq10.7$
is much larger than one, and therefore the time evolution of the
scissors mode is substantially a replica of that of the angular
velocity in the trap. We have observed a similar behavior also for
collision-induced scissors oscillations of a binary BEC system
\cite{2bec}.
\begin{figure}[t]
\centerline{\includegraphics[width=7.5cm,clip=]{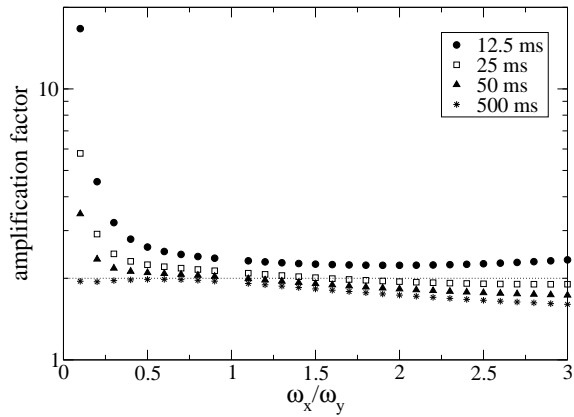}}
\caption{Calculated amplification factor $\sqrt{1+\omega_{sc}^2 F^2}$
 of a scissors mode as a function of $\omega_x/\omega_y$, for various
expansion times: $t_{exp}=25,50,100, 500$~ms. Here the the scissor
frequency is kept fixed to $\omega_{sc}=2\pi\times200.7$~Hz and the
radial and axial frequencies are varied accordingly.}
\label{fig:amp-th}
\end{figure}

To study the role played by the trap geometry in the amplification mechanism,
we have calculated the ratio between the final amplitude and
the trapped one as a function of the trap anisotropy
$\omega_x/\omega_y$, for various expansion times.  In Fig.\
\ref{fig:amp-th} we show the results, obtained by keeping fixed
the scissor frequency to the value $2\pi\times200.7$~Hz, and
varying the radial and axial frequencies accordingly.  The
behavior shown is independent of the initial angle
$\theta_{0i}$, in the small angle regime. This picture confirms
that in cigar-shaped traps, for typical expansion times
($t_{exp}\simeq10\div30$~ms), the scissors mode amplitude is substantially
amplified by the expansion. It also shows that the amplification
factor asymptotically tends to a value $\sim2$, regardless
the trap geometry. It is worth noticing that this asymptotic
behavior is similar to that of a 2D condensate in the limit
$\omega_x/\omega_y=1$, where the relation (\ref{eq:fit}) can be
demonstrated analytically \cite{castindum}. Notice that in
contrast Eq.\ (\ref{eq:thermal}) would not imply any amplification
of the asymptotic oscillations of the thermal cloud.

As a final remark we recall that in general the scissors modes are
coupled to compressional modes, since a rotation of the condensate
can induce shape deformation if the rotation angle is not small
enough \cite{scissors}. In our trap, by increasing the amplitude
of the dipolar oscillation, and hence the angular velocity
acquired by the condensate, it is also possible to access the
non-linear regime, where the scissors mode couples to quadrupole
shape oscillations.

In conclusion, we have investigated the
expansion of the scissors mode of BECs. We have shown that the
asymptotic oscillation is still characterized by a sinusoidal
behavior, which depends linearly on the initial amplitude and
angular velocity, resulting in an enhancement of the final
amplitude. The capability of exciting
scissors oscillations in elongated condensates opens the
possibility of future studies of rotating superfluids in a novel
regime of large moments of inertia and small angular
velocities.\par 

We acknowledge useful discussion with F.~S.~Cataliotti, P.~Pedri, and
S.~Stringari, and the helpful contribution by L.~Fallani. 
This work was supported by MIUR, by ECC under the
Contract HPRICT1999-00111, and by INFM, PRA ``Photonmatter''.


\begin{thebibliography}{}

\bibitem{vortici} M.~R.~Matthews, {\it et al.},
Phys. Rev. Lett. {\bf 83}, 2498 (1999); K.~W.~ Madison,{\it et
al.},
Phys.  Rev. Lett. {\bf 84}, 806 (2000).

\bibitem{ketterle} C.~Raman, {\it et al.},
Phys. Rev. Lett. {\bf 83}, 2502 (1999).

\bibitem{firenze} S. Burger, {\it et al.}, Phys. Rev. Lett. {\bf 86},
4447 (2001).

\bibitem{scissors} D.~Guery-Odelin and S.~Stringari,
Phys. Rev. Lett. {\bf 83}, 4452 (1999).

\bibitem{scissors_foot} O.~M.~Marag\`o, S.~A.~Hopkins, J.~ Artl, 
E.~Hodby, G.~He\-shen\-blai\-kner, and C.~J.~Foot
Phys. Rev. Lett. {\bf 84}, 2056 (2000);
O.~M.~Marag\`o, G.~He\-shen\-blai\-kner, E.~Hodby, S.~A.~Hopkins, 
and C.~J.~Foot, {\tt cond-mat/0109193}.

\bibitem{edwards} M.~Edwards, C.~W.~Clark, P.~Pedri, L.~Pitaevskii, 
and S.~Stringari, Phys. Rev. Lett. {\bf 88}, 070405 (2002).

\bibitem{foot} G. He\-shen\-blai\-kner, E. Hodby, S. A. Hopkins, 
O. M. Marag\`o, and C.~J.~Foot, Phys. Rev. Lett. {\bf 88}, 070406 (2002).

\bibitem{stringari} S.~Stringari, Phys. Rev. Lett. {\bf 77}, 2360 (1996).

\bibitem{BEC_review} F.~Dalfovo, {\it et al.},
Rev. Mod. Phys. {\bf 71}, 463 (1999).

\bibitem{tf} F.~Dalfovo, C.~Minniti, and L.~P.~Pitaevskii,
Phys. Rev. A {\bf 56}, 4855 (1997);
S.~Sinha and Y.~Castin, Phys. Rev. Lett. {\bf 87},
190402 (2001).

\bibitem{science} For more details on the experimental apparatus see:
G. Modugno, G. Ferrari, G. Roati, R. J. Brecha, A. Simoni, and M.
Inguscio, Science {\bf 294}, 1320 (2001).

\bibitem{zambelli} F.~Zambelli and S.~Stringari,
Phys. Rev. A {\bf 63}, 033602 (2001).


\bibitem{2bec} G.~Modugno, M. Modugno, F. Riboli, G. Roati, and M. Inguscio,
preprint {\tt cond-mat/0205485}.

\bibitem{castindum}
Y.~Castin and R.~Dum, Eur. Phys. J. D {\bf 7}, 399 (1999).



\end{thebibliography}
\end{document}